\documentstyle[aps,prd,tighten,epsf]{revtex}
\begin{document}
\draft
\newcommand{\be}{\begin{equation}}
\newcommand{\ee}{\end{equation}}
\newcommand{\ben}{\begin{eqnarray}}
\newcommand{\een}{\end{eqnarray}}

\newcommand{\la}{{\lambda}}
\newcommand{\Om}{{\Omega}}
\newcommand{\ta}{{\tilde a}}
\newcommand{\bg}{{\bar g}}
\newcommand{\bh}{{\bar h}}
\newcommand{\si}{{\sigma}}
\newcommand{\th}{{\theta}}
\newcommand{\C}{{\cal C}}
\newcommand{\D}{{\cal D}}
\newcommand{\cA}{{\cal A}}
\newcommand{\cT}{{\cal T}}
\newcommand{\cO}{{\cal O}}
\newcommand{\eeo}{\cO ({1 \over E})}
\newcommand{\G}{{\cal G}}
\newcommand{\cL}{{\cal L}}
\newcommand{\T}{{\cal T}}
\newcommand{\M}{{\cal M}}

\newcommand{\p}{\partial}
\newcommand{\na}{\nabla}
\newcommand{\ssum}{\sum\limits_{i = 1}^3}
\newcommand{\dssum}{\sum\limits_{i = 1}^2}
\newcommand{\tal}{{\tilde \alpha}}

\newcommand{\tp}{{\tilde \phi}}
\newcommand{\tPhi}{\tilde \Phi}
\newcommand{\tpsi}{\tilde \psi}
\newcommand{\tim}{{\tilde \mu}}
\newcommand{\tr}{{\tilde \rho}}
\newcommand{\tir}{{\tilde r}}
\newcommand{\rp}{r_{+}}
\newcommand{\hr}{{\hat r}}
\newcommand{\rv}{{r_{v}}}
\newcommand{\dr}{{d \over d \hr}}
\newcommand{\dR}{{d \over d R}}

\newcommand{\hhf}{{\hat \phi}}
\newcommand{\hhM}{{\hat M}}
\newcommand{\hhQ}{{\hat Q}}
\newcommand{\hht}{{\hat t}}
\newcommand{\hhr}{{\hat r}}
\newcommand{\hhS}{{\hat \Sigma}}
\newcommand{\hhD}{{\hat \Delta}}
\newcommand{\hhm}{{\hat \mu}}
\newcommand{\hro}{{\hat \rho}}
\newcommand{\hhz}{{\hat z}}

\newcommand{\tD}{{\tilde D}}
\newcommand{\tB}{{\tilde B}}
\newcommand{\tV}{{\tilde V}}
\newcommand{\hT}{\hat T}
\newcommand{\tF}{\tilde F}
\newcommand{\tT}{\tilde T}
\newcommand{\hC}{\hat C}
\newcommand{\ep}{\epsilon}
\newcommand{\bep}{\bar \epsilon}
\newcommand{\ppp}{\varphi}
\newcommand{\Ga}{\Gamma}
\newcommand{\ga}{\gamma}
\newcommand{\hth}{\hat \theta}
\title{Physical Process Version of the First Law of Thermodynamics
for Black Holes in Einstein-Maxwell Axion-dilaton Gravity}

\author{Marek Rogatko}
\address{Institute of Physics \protect \\
Maria Curie-Sklodowska University \protect \\
20-031 Lublin, pl.Marii Curie-Sklodowskiej 1, Poland \protect \\
rogat@tytan.umcs.lublin.pl \protect \\
rogat@kft.umcs.lublin.pl}
\date{\today}
\maketitle
\smallskip
\bigskip
\begin{abstract}
We derive general formulae for the first order variation of the ADM mass,
angular momentum for linear perturbations of a stationary background
in Einstein-Maxwell axion-dilaton gravity being the low-energy limit
of the heterotic string theory. All these variations were expressed in 
terms of the perturbed matter energy momentum tensor and the perturbed 
charge current density. Combining these expressions we reached to the
form of the {\it physical process version} of the first law of black hole dynamics
for the stationary black holes in the considered theory being the strong support
for the cosmic censorship hypothesis.

\end{abstract}
\baselineskip=18pt
\par
\section{Introduction}
Black hole thermodynamics has played a key role in analyzing the 
character of gravity in general and quantum gravity in particular.
The first law of black hole mechanics as derived by 
Bardeen {\it et al.}\cite{bar73} considered a linear perturbations of a stationary, 
electrovac black hole to another stationary electrovac black hole. It bounded
small changes in the mass of a stationary, axisymmetric black hole
to small changes in its horizon area, angular momentum and the
properties of a stationary perfect fluid surrounding it.\\
For an arbitrary diffeomorphism invariant Lagrangian theory with metric
and matter fields possessing stationary and axisymmetric black hole solutions
which were asymptotically flat and had bifurcate Killing horizon the problem was
considered by Wald and collaborators in \cite{wal93,iye94,iye95}. It was revealed 
that the first law of black hole mechanics might be depicted in a form only
involving surface integrals on the sphere at spatial infinity and the 
bifurcation sphere of the black hole horizon. Iyer  \cite{iye97}
applied the aforementioned analysis to the problem of stationary,
axisymmetric black hole surrounded by a perfect fluid.
\par
It seems that the ultimate nature of quantum gravity should of the form of
Lagrangian that describes  the dynamics of classical {\it background fields}
for sufficiently weak fields at sufficiently large distances.
This low-energy effective action may also contain higher curvature 
terms and higher derivative terms in the metric and all other matter fields.
Jacobson {\it et al.}\cite{jac94} computed black hole entropy in a generally covariant
theories including arbitrary higher derivative interactions. In \cite{jac95}
they examined the zeroth and second law of black hole thermodynamics within
context of effective generalized gravitational actions including higher
curvature interactions. They showed that entropy can never decrease for 
quasi-stationary process when black hole acquired positive energy matter.
This is independent of the details of the gravitational actions.
Koga and Maeda \cite{kog98} considered black hole thermodynamics in a generalized
theory of gravity which admitted Lagrangian being an arbitrary function of 
the metric, Ricci tensor and a scalar field. They showed that all thermodynamical
variables defined in \cite{wal93,iye94} are the same in original frame and in the
Einstein frame, under the assumptions that spacetimes in both frames were 
asymptotically flat, regular and possessed event horizons with non-zero 
temperature.
\par
Nowadays there has been an active period of constructing black hole 
solutions in the string theories (see \cite{you99} and references therein).
The low-energy limit of the heterotic string theory compactified on
a six-dimensional torus consists of the pure $N = 4,~d = 4$ supergravity 
coupled to $N = 4$ super Yang-Mills.
The bosonic 
sector of the low-energy limit of the heterotic string theory compactified on a 
six-torus in called Einstein-Maxwell axion-dilaton (EMAD) gravity. This theory provides
a simple framework for studying classical solutions which can be considered
as solutions of the full effective string theory.
In \cite{rog98} the first law of black hole mechanics in EMAD gravity was derived.
This derivation was true for arbitrary perturbation of a stationary
black hole in this theory.\\
It is also possible to consider {\it a physical process} when matter is thrown
into initially stationary black hole. The tantalizing question is whether the
black hole settles down to a final stationary state. As was pointed out if 
this would be violated than this event provided the strong evidence against
cosmic censorship. However if one gets the proof of the {\it a physical process}
version of the first law of thermodynamics and it will be supporting evidence for
the cosmic censorship.
The physical version of the first law of black hole thermodynamics was proved 
in \cite{wal94} and in \cite{gao01} it was generalized to the case of charged black 
holes in  Einstein-Maxwell (EM) theory.
\par
In this paper we shall try to provide some continuity
with our previous work \cite{rog98} and find the {\it physical process version}
of the first law of black hole dynamics in EMAD gravity. In section II
we find the first order variation for the Arnowitt-Deser-Misner
(ADM) mass and angular momentum for the
linear perturbations of a stationary background in EMAD gravity. Then
we established the {\it physical process version} of the first law of black hole dynamics for
stationary, charged black hole solutions in the considered theory. We found
that the form of it is the same as derived in Ref.\cite{rog98}, provided
in this way the strong support for the cosmic censorship hypothesis.
In section III we concluded our results. Our notation and conventions follow that 
which was used in \cite{wal84}.

\section{Physical process version of the first law of black hole mechanics}
In this section we shall consider the effective Lagrangian of the low-energy
heterotic string theory compactified on a six-dimensional torus. The bosonic 
sector of this theory with a simple vector field is called EMAD gravity.
This theory provides a non-trivial generalization of the ordinary EM gravity.
It consists of a coupled system containing a metric $g_{\mu \nu}$,
$U(1)$ vector fields $A_{\mu}$, a dilaton field $\phi$ and three-index
antisymmetric tensor field, namely \cite{sha91}
\be
{\bf L } = {\bf \ep} \bigg(
R - 2 (\na \phi )^2 - {1 \over 3} e^{- 4 \phi} H_{\alpha \beta \ga}H^{\alpha \beta \ga}
- e^{- 2 \phi}F_{\alpha \beta} F^{\alpha \beta} \bigg),
\label{ll}
\ee
where by $ {\bf \ep}$ we denoted the volume element,
$F_{\mu \nu} = 2 \na_{[\mu}A_{\nu]}$ and $H_{\alpha \beta \ga}$ stands for
three-index antisymmetric tensor field defined by
\be
H_{\alpha \beta \ga} = \na_{\alpha} B_{\beta \ga} - A_{\alpha} F_{\beta \ga} +
cyclic.
\ee
We proceed now to finding the first order variation of the conserved quantities
in the theory under consideration. Our main task will be to obtain the explicit 
formulae for the variation of mass and angular momentum in EMAD gravity.
Thus, calculating the first order variation of the Lagrangian (\ref{ll}), one finds
\ben \label{dl}
\delta {\bf L} &=& {\bf \epsilon} \bigg(
G_{\mu \nu} - T_{\mu \nu}(\phi, F, H) \bigg)~ \delta g^{\mu \nu}
+ 4 \bigg( \na_{\mu}(e^{- 2 \phi} F^{\mu \nu}) + e^{- 4 \phi} 
H^{\alpha \beta \nu} F_{\alpha \beta} \bigg)~ \delta A_{\nu} \\ \nonumber
&+& 2 \na_{\mu} (e^{- 4 \phi} H^{\mu \alpha \beta})~ \delta B_{\alpha \beta} + 
4 \bigg( \na^{2} \phi + {1 \over 3} e^{- 4 \phi} H^{2} 
+ {1 \over 2} e^{- 2 \phi}F^{2} \bigg)~ \delta \phi + d {\bf \Theta},
\een
where the energy momentum tensor is given by
\be
T_{\mu \nu}(\phi, F, H) =
2 \na_{\mu} \phi \na_{\nu} \phi - g_{\mu \nu} \big( \na \phi \big)^2 +
e ^{- 2 \phi} \bigg( 
2 F_{\mu \ga}F_{\nu}{}{}^{\ga} - {1 \over 2} g_{\mu \nu}F^2 \bigg) +
e^{- 4 \phi} \bigg(
H_{\mu \alpha \beta} H_{\nu}{}{}^{\alpha \beta} - 
{1 \over 6} g_{\mu \nu} H^{2} \bigg).
\ee
The totally divergent term in (\ref{dl}) is a functional of the field variables
$(A_{\mu}, B_{\mu \nu}, \phi)$ and their variations $(\delta A_{\mu},
\delta B_{\mu \nu}, \delta \phi)$, which for simplicity we denote respectively 
by $\psi_{\alpha}$ and $\delta \psi_{\alpha}$. From equation (\ref{dl}) we have the following
expression for the symplectic three-form
$\Theta_{\alpha \beta \ga}[\psi_{\alpha}, \delta \psi_{\alpha}]$
\be
\Theta_{\alpha \beta \ga}\big[\psi_{\alpha}, \delta \psi_{\alpha} \big] =
\ep_{\mu \alpha \beta \ga} v^{\mu},
\ee
where
\be
v^{\mu} = 
\omega^{\mu} - 4 e^{- 2 \phi}F^{\mu \nu}~ \delta A_{\nu} -
2 e ^{-4 \phi} H^{\mu \alpha \beta}~ \delta B_{\alpha \beta} +
4 e^{- 4 \phi} H^{\ga \mu \beta} A_{\ga}~ \delta A_{\beta} - 
4 \na^{\mu} \phi~ \delta \phi,
\ee
and
\be
\omega_{\mu} = \na^{\alpha} \delta g_{\alpha \mu} - \na_{\mu} 
\delta g_{\beta}{}{}^{\beta}.
\ee
The inspection of (\ref{dl}) enables one to read off
the source-free EMAD Eqs. of motion. They  yield
\ben
G_{\mu \nu} - T_{\mu \nu}(\phi, F, H) &=& 0, \\
\na_{\nu}(e^{- 2 \phi} F^{\mu \nu}) + e^{- 4 \phi} 
H^{\alpha \beta \mu} F_{\alpha \beta} &=& 0,\\
\na_{\mu} (e^{- 4 \phi} H^{\mu \alpha \beta}) &=& 0,\\
 \na^{2} \phi + {1 \over 3} e^{- 4 \phi} H^{2} 
+ {1 \over 2} e^{- 2 \phi}F^{2} &=& 0.
\een
When one identifies the variations of the fields $\delta \psi_{\alpha}$
with a general coordinate transformations ${\cal L}_{\xi} \psi_{\alpha}$
induced by an arbitrary Killing vector $\xi_{\alpha}$,
the Noether three-form with respect to this Killing vector $\xi_{\alpha}$ implies
\cite{iye94,iye95}
\be
{\cal J}_{\alpha \beta \ga} = \ep_{\rho \alpha \beta \ga} {\cal J}^{\rho}
\big[\psi_{\alpha}, {\cal L}_{\xi} \psi_{\alpha}\big],
\ee
where the vector field ${\cal J}^{\rho}
\big[ \psi_{\alpha}, {\cal L}_{\xi} \psi_{\alpha}\big]$ is given by
\be
{\cal J}^{\delta}
\big[\psi_{\alpha}, {\cal L}_{\xi} \psi_{\alpha}\big] = \Theta^{\delta}
\big[\psi_{\alpha}, {\cal L}_{\xi} \psi_{\alpha}\big] - \xi^{\delta} L.
\label{curr}
\ee
From relation (\ref{curr}) we can establish
the resultant expression for the Noether
current three-form with respect to the Killing vector field $\xi_{\alpha}$, namely
\ben \label{ttt}
{\cal J}_{\alpha \beta \ga} &=&
d Q_{\alpha \beta \ga}^{GR} + 2 \ep_{\delta \alpha \beta \ga}
\bigg(
G^{\delta}{}{}_{\eta} - T^{\delta}{}{}_{\eta}(\phi, H, F)
\bigg) \xi^{\eta} - 4 \ep_{\delta \alpha \beta \ga}
\bigg[ \na_{\mu}(e^{-2 \phi} F^{\delta \mu}) + e^{- 4 \phi} H^{\delta \mu \nu}
F_{\mu \nu} \bigg]
\xi^{\rho} A_{\rho} \\ \nonumber
&-& 4 \ep_{\delta \alpha \beta \ga} \na_{\mu}(e^{- 4 \phi} H^{\delta \mu \nu}) B_{\nu \eta}
\xi^{\eta} - 4 \ep_{\delta \alpha \beta \ga} \na_{\mu}(e^{- 4 \phi} H^{\rho \delta \nu})
A_{\rho} \xi^{\eta}A_{\eta} \\ \nonumber
&-& 4 \ep_{\delta \alpha \beta \ga}\na_{\mu}
\bigg( e^{- 2 \phi} F^{\delta \mu} \xi^{\rho} A_{\rho} \bigg)
+ 4 \ep_{\delta \alpha \beta \ga} \na_{\mu}
\bigg( e^{- 4 \phi} H^{\delta \mu \nu} \xi^{\rho}B_{\nu \rho} \bigg)\\ \nonumber
&+& 4 \ep_{\delta \alpha \beta \ga} \na_{\mu}
\bigg( e^{- 4 \phi}H^{\rho \delta \mu} A_{\rho} \xi^{\eta} A_{\eta} \bigg),
\een
where we have denoted by
$Q_{\alpha \beta \ga}^{GR} = - \ep_{\alpha \beta \ga \delta}\na^{\ga} \xi^{\delta}$.
Using differentiation by parts we obtain the following expression for 
equation (\ref{ttt}):
\ben
{\cal J}_{\alpha \beta \ga} &=&
d Q_{\alpha \beta \ga} +
2 \ep_{\delta \alpha \beta \ga}
\bigg(
G^{\delta}{}{}_{\eta} - T^{\delta}{}{}_{\eta}(\phi, H, F)
\bigg) \xi^{\eta}
- 4 \ep_{\delta \alpha \beta \ga}
\bigg[ \na_{\mu}(e^{-2 \phi} F^{\delta \mu}) + e^{- 4 \phi} H^{\delta \mu \nu}
F_{\mu \nu} \bigg]
\xi^{\rho} A_{\rho} \\ \nonumber
&-& 4 \ep_{\delta \alpha \beta \ga} \na_{\mu}(e^{- 4 \phi} H^{\delta \mu \nu}) B_{\nu \eta}
\xi^{\eta} - 4 \ep_{\delta \alpha \beta \ga} \na_{\mu}(e^{- 4 \phi} H^{\rho \delta \mu})
A_{\rho} \xi^{\eta}A_{\eta}.
\een
Having in mind \cite{iye95} that ${\cal J}[\xi] = d Q[\xi] + \xi^{\alpha}{\bf C_{\alpha}}$,
where  ${\bf C_{\alpha}}$ is an $(n - 1)$ form locally constructed from
the dynamical fields we may identify $Q_{\alpha \beta }$
as the Noether charge. The result yields
\ben
Q_{\alpha \beta } &=&  Q_{\alpha \beta }^{GR} +  Q_{\alpha \beta }{}{}^{(F)}
+ Q_{\alpha \beta }{}{}^{(H)} +  Q_{\alpha \beta}{}{}^{(A-H)} = \\ \nonumber
&-& \ep_{\alpha \beta \ga \delta} \na^{\ga} \xi^{\delta}
- \ep_{\alpha \beta \ga \delta} e^{- 2 \phi} F^{\ga \delta} A_{\rho} \xi^{\rho}
+ \ep_{\alpha \beta \ga \delta} e^{- 4 \phi} 
H^{\ga \delta \eta} B_{\eta \rho} \xi^{\rho} +
\ep_{\alpha \beta \ga \delta} e^{- 4 \phi} 
H^{\ga \delta \eta} A_{\eta } A_{\rho} \xi^{\rho},
\een
and also 
 $C_{\alpha \beta \ga \mu}$
 may be read off the above expression. It is given by the following formula:
\ben
C_{\alpha \beta \ga \mu } &=& 2 \ep_{\delta \alpha \beta \ga}
\bigg(
G^{\delta}{}{}_{\mu} - T^{\delta}{}{}_{\mu}(\phi, H, F)
\bigg)
- 4 \ep_{\delta \alpha \beta \ga}
\bigg[ \na_{\rho}(e^{-2 \phi} F^{\delta \rho}) + e^{- 4 \phi} H^{\delta \ga \nu}
F_{\ga \nu} \bigg]
\xi^{\rho} A_{\mu} \\ \nonumber
&-& 4 \ep_{\delta \alpha \beta \ga} \na_{\rho}(e^{- 4 \phi} H^{\delta \rho \nu}) B_{\nu \mu}
- 4 \ep_{\delta \alpha \beta \ga} \na_{\mu}(e^{- 4 \phi} H^{\rho \delta \mu})
A_{\rho} A_{\mu}.
\een
When the quantity ${\bf C_{\alpha}} = 0$, we have equations of motion 
fullfiled. On the other hand,
in the case 
when it does not hold, one gets
\ben \label{eqnm}
G_{\mu \nu} - T_{\mu \nu}(\phi, F, H) &=& T_{\mu \nu}(matter), \\ \label{a2}
\na_{\nu}(e^{- 2 \phi} F^{\mu \nu}) + e^{- 4 \phi} 
H^{\alpha \beta \mu} F_{\alpha \beta} &=& - j^{\mu}(matter),\\ \label{a3}
\na_{\nu} (e^{- 4 \phi} H^{\nu \alpha \beta}) &=& 0,\\ \label{a4}
 \na^{2} \phi + {1 \over 3} e^{- 4 \phi} H^{2} 
+ {1 \over 2} e^{- 2 \phi}F^{2} &=& 0.
\label{a5}
\een
Let us have that $(g_{\mu \nu}, A_{\mu}, B_{\mu \nu}, \phi)$ be the solution of the 
source-free EMAD equations of motion and let us assume that 
$(\delta g_{\mu \nu}, \delta A_{\mu}, \delta B_{\mu \nu}, \delta \phi)$
be linearized perturbations fulfilling EMAD gravity equations with
sources $\delta T_{\mu \nu}(matter)$ and $\delta j_{\mu}(matter)$.
It implies that for a perturbed $\delta C_{\alpha \beta \ga \mu }$
quantity we arrive at the following relation:
\be
\delta C_{\alpha \beta \ga \mu } = \ep_{\delta \alpha \beta \ga}
\bigg[ 2 \delta T^{\delta}{}{}_{\mu}(matter) + 
4 A_{\mu} \delta j^{\delta}(matter) \bigg].
\label{ccc}
\ee
Equation (\ref{ccc}) enables us to 
obtain the explicit formula 
for the variation of the conserved quantity
$\delta H_{\xi}$ associated with the Killing vector field \cite{wal00}.
It satisfies the following:
\be
\delta H_{\xi} = \int_{\Sigma} \bigg[
2 \xi^{\mu}\delta T^{\delta}{}{}_{\mu}(matter) + 4 \xi^{\mu}
A_{\mu} \delta j^{\delta}(matter) \bigg] +
\int_{\p \Sigma}\bigg[
\delta Q(\xi) - \xi \cdot \Theta \bigg].
\label{hh}
\ee
Choosing $\xi^{\alpha}$ to be an asymptotic time translation $t^{\alpha}$
we write $M = H_{t}$ and get the desired variation of the ADM mass
\be
\delta M = \int_{\Sigma} \bigg[
2 t^{\mu}\delta T^{\delta}{}{}_{\mu}(matter) + 4 t^{\mu}
A_{\mu} \delta j^{\delta}(matter) \bigg] +
\int_{\p \Sigma}\bigg[
\delta Q(t) - t \cdot \Theta \bigg].
\label{mm}
\ee
On the other hand, when one chooses $\xi^{\alpha}$ to be an asymptotic  rotation
$\varphi^{\alpha}$ we have the variation of angular momentum
\be
\delta J = \int_{\Sigma} \bigg[
2 \varphi^{\mu}\delta T^{\delta}{}{}_{\mu}(matter) + 4 \varphi^{\mu}
A_{\mu} \delta j^{\delta}(matter) \bigg] +
\int_{\p \Sigma}\bigg[
\delta Q(\varphi) - \varphi \cdot \Theta \bigg].
\label{jj}
\ee
Now, we proceed to {\it the physical version} of the first law of black hole
mechanics. 
We shall consider a classical black hole solution to EMAD gravity theory which
satisfies the equations of motion (\ref{eqnm}-\ref{a5}). The perturbation of the black hole
is gained by dropping matter into it. We assume further that the black hole 
settles down to a stationary final state and its change of mass, angular momentum
and changes of currents of the fields in the theory under
consideration can be found.
Let $(g_{\alpha \beta}, A_{\alpha}, B_{\alpha \beta}, \phi)$
be a solution of the source-free EMAD gravity equations of motion
corresponding to a stationary black hole. The Killing vector field
normal to the horizon
is given as follows $\xi_{\alpha} = t_{\alpha} + \Omega \varphi_{\alpha}$,
where $\Omega$ is the angular velocity of the black hole.
\\
As in \cite{gao01} we suppose that $\Sigma_{0}$ is an asymptotically flat
hypersurface which terminates on the event horizon. We shall consider
the initial data on $\Sigma_{0}$ for a linearized perturbations
$(\delta g_{\mu \nu}, \delta A_{\mu}, \delta B_{\mu \nu}, \delta \phi)$
with $\delta T_{\mu \nu}(matter)$ and $\delta j^{\mu}(matter)$. We 
require that $\delta T_{\mu \nu}(matter)$ and $\delta j^{\mu}(matter)$ will
vanish at infinity and the initial data for 
$(\delta g_{\mu \nu}, \delta A_{\mu}, \delta B_{\mu \nu}, \delta \phi)$
vanish in the vicinity of the black hole horizon $\cal H$ on 
the hypersurface $\Sigma_{0}$. It envisages the fact that for the initial time
$\Sigma_{0}$, the considered black hole is unperturbed. In our considerations
we assume that all the matter, charges and so on will fall into black hole
and the black hole will settle down to another stationary black hole solution
of the source-free equations of EMAD gravity.\\
Our main aim is to compute $\delta M,~ \delta J,~ \delta Q$ and $\delta A$ for the final
state of black hole and verify the validity of the first law of black hole mechanics
in EMAD gravity.
From relations (\ref{mm}) and (\ref{jj}), one has
\be
\delta M - \Omega \delta J = - \int_{\Sigma_{0}} \ep_{\delta \alpha \beta \ga}
\bigg[
2 \xi^{\mu}\delta T^{\delta}{}{}_{\mu}(matter) + 4 \xi^{\mu}
A_{\mu} \delta j^{\delta}(matter) \bigg] =
 \int_{\Sigma_{0}} \alpha^{\delta} n_{\delta} \bep_{\alpha \beta \ga},
\label{war}
\ee
where $n_{\mu}$ is the future directed unit normal vector to the hypersurface
$\Sigma_{0}$ and $\bep_{\alpha \beta \ga} = n^{\mu} \ep_{\mu \alpha \beta \ga}$.
The fact that our assumption is that all the matter fall into black hole 
enables one to replace $n_{\mu}$ by $k_{\mu}$ which is tangent vector to null 
geodesic generators of the event horizon 
$\cal H$ of the black hole in equation (\ref{war}) and $\cal H$ for $\Sigma_{0}$.\\
The second term in $\alpha^{\delta}$ can be written in the form
\be
ST = - 4 \int_{\cal H} \Phi_{BH}~\delta j^{\mu} k_{\mu}\bep_{\alpha \beta \ga},
\ee
where we have denoted by $\Phi_{BH} = - (\xi^{\mu}A_{\mu})\mid_{\cal H}$.
Using the Raychanduri's equation (see, e.g., \cite{wal84})
and because of the symmetry
of the background we have ${\cal L}_{\xi} A_{\mu} = 0$, one can show that
$\Phi_{BH}$ is constant on black hole horizon $\cal H$.
By $\delta Q(matter)$ we denote the flux of the charge into the black hole
\be
\delta Q(matter) = - 4 \int_{\cal H} \delta j^{\mu}(matter) k_{\mu} \bep_{\alpha \beta \ga}.
\ee
Consequently having in mind equation
(\ref{a2}) and the fact that $H_{\alpha \beta \ga}k^{\alpha} k^{\beta} = 0$
which implies that $H_{\alpha \beta \ga}k^{\alpha} \propto k^{\beta}$ and hence
the pull-back to $\cal H$ of $H_{\alpha \beta \ga}k^{\alpha}$ vanishes, we obtain
only the dependence on the dilaton field $\phi$ and $F_{\mu \nu}$.
Therefore the expression for the flux of the charge implies
\be
\delta Q(matter) = - 4 \int_{\cal H}\bigg[
\na_{\mu} (e^{- 2 \phi} F^{\nu \mu}) k_{\nu}\bep_{\alpha \beta \ga}
\bigg] = \delta Q_{(\phi-F)}.
\ee
Substituting it into relation (\ref{war}) we conclude the following:
\be
\delta M - \Omega \delta J - \Phi_{BH}~\delta Q_{(\phi-F)} =
2 \int_{\cal H} \delta T^{\mu}{}{}_{\nu}~\xi^{\nu}~k_{\mu}.
\ee
In order to find the change in the area of the black hole we use the fact that
null generators of the black hole horizon of the perturbed black hole
coincide with the null generators of the unperturbed
stationary black hole $\delta k_{\mu} \propto k_{\mu}$ and since the
expansion $\theta$
and shear $\sigma_{\mu \nu}$ vanish in the stationary background, the perturbed 
Raychanduri's equation is of the form
\be
{d(\delta \theta) \over d \lambda} =
- \delta \bigg(
T_{\mu \nu}(total) k^{\mu} k^{\nu} \bigg) \mid_{\cal H} =
- \delta \bigg(
T_{\mu \nu}(matter) \bigg)
 k^{\mu} k^{\nu}\mid_{\cal H} 
- \delta \bigg(
T_{\mu \nu}(\phi, F, H) \bigg) 
k^{\mu} k^{\nu} \mid_{\cal H},
\label{ray}
\ee 
where we exploit the fact that 
$T(\phi, F, H)_{\mu \nu} k^{\mu}k^{\nu} \mid_{\cal H} = 0$ and $\delta k_{\mu}
\propto k_{\mu}$ to eliminate terms in the form
$T(\phi, F, H)_{\mu \nu} k^{\mu} \delta k^{\nu}$.
But we have left with the terms of the form 
$\delta
T_{\mu \nu}(\phi, F, H)
k^{\mu} k^{\nu} \mid_{\cal H}$. We shall compute them part by part
\be
\delta
T_{\mu \nu}(F)
k^{\mu} k^{\nu} \mid_{\cal H} =
\bigg(
4 \delta
F_{\mu \ga}F_{\nu}{}{}^{\ga} - {1 \over 2} \delta g_{\mu \nu}F^2 - g_{\mu \nu}
F_{\alpha \beta} \delta F^{\alpha \beta} \bigg) e^{-2 \phi}k^{\mu}k^{\nu}
+ \delta (e^{-2 \phi}) T_{\mu \nu}(F)k^{\mu}k^{\nu},
\label{ffff}
\ee
where the last two terms in the first part of Eq.(\ref{ffff}) are equal to zero 
because of the fact that $k_{\mu}$ is a null vector in the perturbed as well as in the
unperturbed metric. 
Moreover,
we have that $F_{\mu \nu}k^{\nu} \propto k_{\mu}$, thus because of the
antisymmetry of $\delta F_{\mu \nu}$ this term is equal to zero. The second part
of equation (\ref{ffff}) is also equal to zero as we have seen above.\\
In the case of the energy momentum tensor $T_{\mu \nu}(H)$ for
the three-index antisymmetric tensor field we arrive at the expression
\be 
\delta
T_{\mu \nu}(H)
k^{\mu} k^{\nu} \mid_{\cal H} =
\bigg(
2 \delta H_{\mu \alpha \beta} H_{\nu}{}{}^{\alpha \beta} - 
{1 \over 6} \delta g_{\mu \nu} H^{2} - {1 \over 3}g_{\mu \nu}
\delta H_{\alpha \beta \gamma} H^{\alpha \beta \gamma}
\bigg)e^{- 4 \phi}k^{\mu}k^{\nu} +
\delta (e^{- 4 \phi})T_{\mu \nu}(H) k^{\mu}k^{\nu}.
\label{hhh}
\ee
The previous arguments can be repeated, i.e.,
using the antisymmetry of $\delta H_{\alpha \beta \ga}$ and the property of the vector
field $k_{\mu}$ we reach to the conclusion that the right-hand side of (\ref{hhh}) is
equal to zero.
On the other hand for the dilaton field energy momentum tensor we have
\be 
\delta
T_{\mu \nu}(\phi)
k^{\mu} k^{\nu} \mid_{\cal H} =
\bigg(
4 \delta (\na_{\mu} \phi) \na_{\nu} \phi - \delta g_{\mu \nu} 
\big( \na \phi \big)^2 - 2 g_{\mu \nu} \delta (\na_{\gamma}\phi)
\na^{\gamma} \phi \bigg) k^{\mu}k^{\nu} = 0,
\label{ff}
\ee
where in this case we used the fact that ${\cal L}_{k} \phi = 0$,
as well as the property of $k_{\mu}$ vector field.
Thus, we finally receive from equation (\ref{ray}) only one term which yields
\be
{d(\delta \theta) \over d \lambda} = - \delta
T_{\mu \nu}(matter)
 k^{\mu} k^{\nu}\mid_{\cal H},
\label{aaaa}
\ee
In order to compute Eq.(\ref{aaaa}) one can repeat the same steps as done 
in \cite{wal94}.
Now we shall briefly review the key
points of the procedure
(see also for particulars \cite{wal84,kay91}).
Namely, one begins with integrating the right-hand side of equation (\ref{aaaa}) 
over the horizon of black hole, the changes in the black hole geometry may 
be neglected. Thus, it is possible to substitute for $k_{\mu}$
the following expression 
\be
k_{\mu} = \bigg(
{\p \over \p V} \bigg)_{\mu} = {1 \over \kappa V} \bigg(
t_{\mu} + \Omega \varphi_{\mu} \bigg),
\ee
where $\kappa$ is the surface gravity, $V$ is
an affine parameter
along the null geodesics tangent to $\xi_{\beta}$
which generates the adequate Killing horizon.
One should remark \cite{wal94} 
that if the function $v$ ( it is called {\it Killing parameter time})
on the portion of Killing horizon
satisfies
$\xi^{\beta} \na_{\beta} v = 1$, then it is related with $V$ by
the expression $V = exp(\kappa v)$.
Next, one multiplies  both sides of the resulting equation by $\kappa V$
and integrates over the horizon.\\
We also recall that $\theta$ measures the local rate of change of 
the cross-sectional area as the observer moves up the null geodesics, i.e.,
$\theta = {1 \over A}{d A \over d \lambda}$, where $\lambda$ is an
affine parameter which parametrized null geodesics generators of the 
horizon. 
The left-hand side of equation (\ref{aaaa}) is evaluated by integration by parts, 
having in mind that $V = 0$ at the lower limit and $\theta$ has to vanish
faster than $1/V$ as the affine parameter tends to infinity.  
The consequence of the above establishes the result
\be
\kappa~ \delta A = \int_{\cal H} \delta
T^{\mu}{}{}_{\nu}(matter) \xi^{\nu} k_{\mu}
\ee
Then, after rescaling the area
we received {\it the physical process version} of the first law of black hole
mechanics in EMAD gravity. It has the form known from \cite{rog98}
being a strong support for the cosmic censorship hypothesis, namely
\be
\delta M - \Omega \delta J - \Phi_{BH}~\delta Q_{(\phi-F)} =
{\kappa \over 8 \pi} ~\delta {\cal A}.
\ee

\section{Conclusions}
In our work we have studied stationary black hole solution to EMAD
gravity being the low-energy limit of the heterotic string theory.
We perturbed the considered black hole by dropping into it some matter. 
Assuming that our black hole will not be destroyed in this process
and eventually settles down to a final stationary state we calculate the change of
the mass, angular momentum, dilaton-gauge field current and change in the
area of the black hole horizon. We obtained in such way the
{\it physical process version} of the first black hole dynamics in EMAD gravity.
It happens that it has the same form as the first law of black hole dynamics
as derived in \cite{rog98}. Then, the proof of 
the {\it physical process version} of the first law of black hole mechanics in
EMAD gravity provides
the strong support
for the cosmic censorship hypothesis in this theory.

\vspace{2cm}
\noindent
{\bf Acknowledgements:}\\
MR was supported in part by KBN grant 5 PO3B 009 21.

\end{document}